\documentclass[a4paper,8pt,twocolumn]{extarticle}

\usepackage{graphics}
\usepackage{epsfig}
\usepackage{amsmath,amssymb,amsthm,bbm}

\renewcommand{\vec}[1]{\mathbf{#1}}
\newcommand{\abs}[1]{\left\vert #1\right\vert}
\newcommand{\ind}[1]{\mathbbm{1}_{#1}}
\newcommand{\intl}{\int\limits}
\newcommand{\R}{\mathbb{R}}
\newcommand{\Tmax}{T_\textup{max}}
\newcommand{\unit}[1]{\,\textup{#1}}
\newcommand{\vd}{\vec{v}_\text{des}}

\theoremstyle{remark}\newtheorem{remark}{Remark}

\graphicspath{{./}{figure/}}

\title{\LARGE\bf How Can Macroscopic Models Reveal \\ Self-Organization in Traffic Flow?}

\author{Emiliano Cristiani\thanks{E. Cristiani is with Istituto per le Applicazioni del Calcolo ``M. Picone'', Consiglio Nazionale delle Ricerche, Rome, Italy {\tt\small e.cristiani@iac.cnr.it}},
		Benedetto Piccoli\thanks{B. Piccoli is with the Department of Mathematical Sciences, Rutgers University - Camden, Camden NJ, USA {\tt\small piccoli@camden.rutgers.edu}},
		Andrea Tosin\thanks{A. Tosin is with Istituto per le Applicazioni del Calcolo ``M. Picone'', Consiglio Nazionale delle Ricerche, Rome, Italy {\tt\small a.tosin@iac.cnr.it}}
}
\date{}

\begin{document}
\maketitle
\thispagestyle{empty}
\pagestyle{empty}

\begin{abstract}
In this paper we propose a new modeling technique for vehicular traffic flow, designed for capturing at a macroscopic level some effects, due to the microscopic granularity of the flow of cars, which would be lost with a purely continuous approach. The starting point is a multiscale method for pedestrian modeling, recently introduced in \cite{cristiani2011mmg}, in which measure-theoretic tools are used to manage the microscopic and the macroscopic scales under a unique framework. In the resulting coupled model the two scales coexist and share information, in the sense that the same system is simultaneously described from both a discrete (microscopic) and a continuous (macroscopic) perspective. This way it is possible to perform numerical simulations in which the single trajectories and the average density of the moving agents affect each other. Such a method is here revisited in order to deal with multi-population traffic flow on networks. For illustrative purposes, we focus on the simple case of the intersection of two roads. By exploiting one of the main features of the multiscale method, namely its dimension-independence, we treat one-dimensional roads and two-dimensional junctions in a natural way, without referring to classical network theory. Furthermore, thanks to the coupling between the microscopic and the macroscopic scales, we model the continuous flow of cars without losing the right amount of granularity, which characterizes the real physical system and triggers self-organization effects, such as, for example, the oscillatory patterns visible at jammed uncontrolled crossroads.
\end{abstract}

\section{Introduction}
Traffic flow along a road is ruled by complex dynamics, which involve mutual interactions among the vehicles. The effect of such interactions is a modification of the traveling velocity of cars through successive slowing down and acceleration maneuvers, resulting globally in self-organized flow patterns that can be clearly seen at large scales. The challenging goal of mathematical models of road traffic should be devising methods able to reveal the spontaneous emergence of macroscopic self-organization from the microscopic interaction dynamics. Indeed, from an engineering point of view, the macroscopic information is ultimately the most useful one for quantitative purposes, since it is generally more robust, i.e., less prone to fluctuations, than the microscopic one, and furthermore it refers to quantities, such as car density and flux, directly measurable in practice. On the other hand, the actual physics of vehicular traffic pertains to the scale of single cars and may not be fully caught simply by averaged approaches, because the most interesting phenomena occur out of equilibrium.

Mathematical models of vehicular traffic have been traditionally focusing on just either scale of description. At the \emph{microscopic} scale, see e.g., \cite{chakroborty2004mmd,gazis1961nfl,helbing2001trs,hoogendoorn2001soa,kerner2002mmp,treiber2000cts}, cars are modeled individually, usually as points, and their interconnected dynamics are formalized through systems of ordinary differential equations inspired by the classical framework of Newtonian mechanics:
$$ \ddot{x}_i=a_i(t,\,x_1,\,\dots,\,x_N,\,\dot{x}_1,\,\dots,\,\dot{x}_N), \qquad i=1,\,\dots,\,N, $$
where $x_i=x_i(t)$ is the position of the $i$-th car at time $t$ along a one-dimensional road and $a_i$ is a material model of its acceleration, which depends on the current position and speed of other neighboring cars (in most cases, of the head car only). Conversely, at the \emph{macroscopic} scale, see e.g., \cite{aw2000rso,bellomo2011mtc,garavello2006tfn,lighthill1955kw2,piccoli2009vtr,richards1956swh}, cars are assimilated to a continuum with density $\rho$ obeying the conservation law:
\begin{equation}
	\frac{\partial\rho}{\partial t}+\frac{\partial}{\partial x}(\rho v)=0,
	\label{eq:cons.law.macro}
\end{equation}
which expresses the conservation in time of the total number of cars, $v$ being their average speed. If \eqref{eq:cons.law.macro} is supplemented by a phenomenological relationship $v=v(\rho)$ linking the speed to the local density (\emph{speed diagram}), one obtains the so-called \emph{first order} models. If, instead, a further equation is joined to \eqref{eq:cons.law.macro}, expressing the balance of linear momentum:
$$ \frac{\partial v}{\partial t}+v\frac{\partial v}{\partial x}=a(\rho,\,v,\,\partial_x\rho,\,\partial_xv), $$
$a$ being now a material model for the average acceleration of cars, one obtains the so-called \emph{second order} models.

The aim of the present paper is to go beyond the just recalled dichotomy microscopic/macroscopic, taking a point of view centered around the concept of vehicular traffic as a complex system. Namely, a system in which the causes leading to macroscopic observable outcomes have to be sought at smaller scales, particularly the microscopic scale of the interactions among vehicles. In more detail, our interest is in the possibility to reproduce, on the macroscopic flow, some self-organized effects typical of car-to-car interactions, such as, for instance, the \emph{traffic light effect}. By this we mean the oscillatory pattern showing up when groups of cars from incoming roads occupy alternately a junction, giving rise to a sequence of density wave packets along the outgoing roads. This usually occurs at jammed crossroads not regulated by any imposed red-green traffic light cycle \cite{helbing2007son}.

In order to achieve our goal, we will evolve a multiscale method, based on the mathematical measure theory, recently proposed for pedestrian traffic modeling \cite{cristiani2011mmg}. The key idea of the method is to describe the distribution in space of a population of moving agents by a measure, which can be understood as their mass or, alternatively, their probability distribution if properly rescaled with respect to the total number of agents. According to the spatial structure of the measure, from such an abstract setting one can derive either a Lagrangian microscopic or an Eulerian macroscopic model: the former is obtained by a combination of Dirac delta's centered at the spatial positions of the agents, the latter by a density making the probability absolutely continuous with respect to the Lebesgue measure. Furthermore, considering that a measure can feature both a singular and a regular part at the same time, one can also obtain multiscale models in which the individual discrete/microscopic and continuous/macroscopic models coexist and automatically complement each other, \emph{without requiring a domain decomposition and transmission conditions between the scales at the interfaces} as it happens instead in most multiscale models available in the literature (see e.g., \cite{helbing2002mms,herty2009mky,lattanzio2010cmm}).

The evolution of our method, that we present in this paper, is twofold:
\begin{itemize}
\item we complement the multiscale coupling with a \emph{multidimensional} coupling. Indeed, we model microscopic dynamics as genuinely two-dimensional, in order to retain the detail of possible lateral displacements of single cars, and the macroscopic dynamics as one-dimensional, for such a detail is actually not needed in a gross average context. This also avoids the nontrivial imposition of boundary conditions along the lateral edges of the roads;
\item we perform a \emph{variable-in-space} multiscale coupling by means of measures parameterized by the space variable, in order to account for the proper scale in the proper portion of the domain.
\end{itemize}

With reference to the aforementioned case study of the traffic light effect, we demonstrate with our model that a self-organized flow cannot be reproduced by purely macroscopic dynamics. The multiscale coupling, with the microscopic scale possibly acting at the junction only, allows instead the macroscopic scale to inherit the right amount of granularity, which generates clearly visible density wave packets beyond the junction.

\section{The multiscale model}
In this section we derive our multiscale model, which is methodologically inspired by the one we developed in \cite{cristiani2011mmg} for crowd dynamics. We first present it in full generality, considering a genuinely two-dimensional domain, then we specialize it to the coupling of one-dimensional macroscopic and two-dimensional microscopic dynamics.

\begin{figure}[!t]
\centering
\includegraphics[scale=0.2]{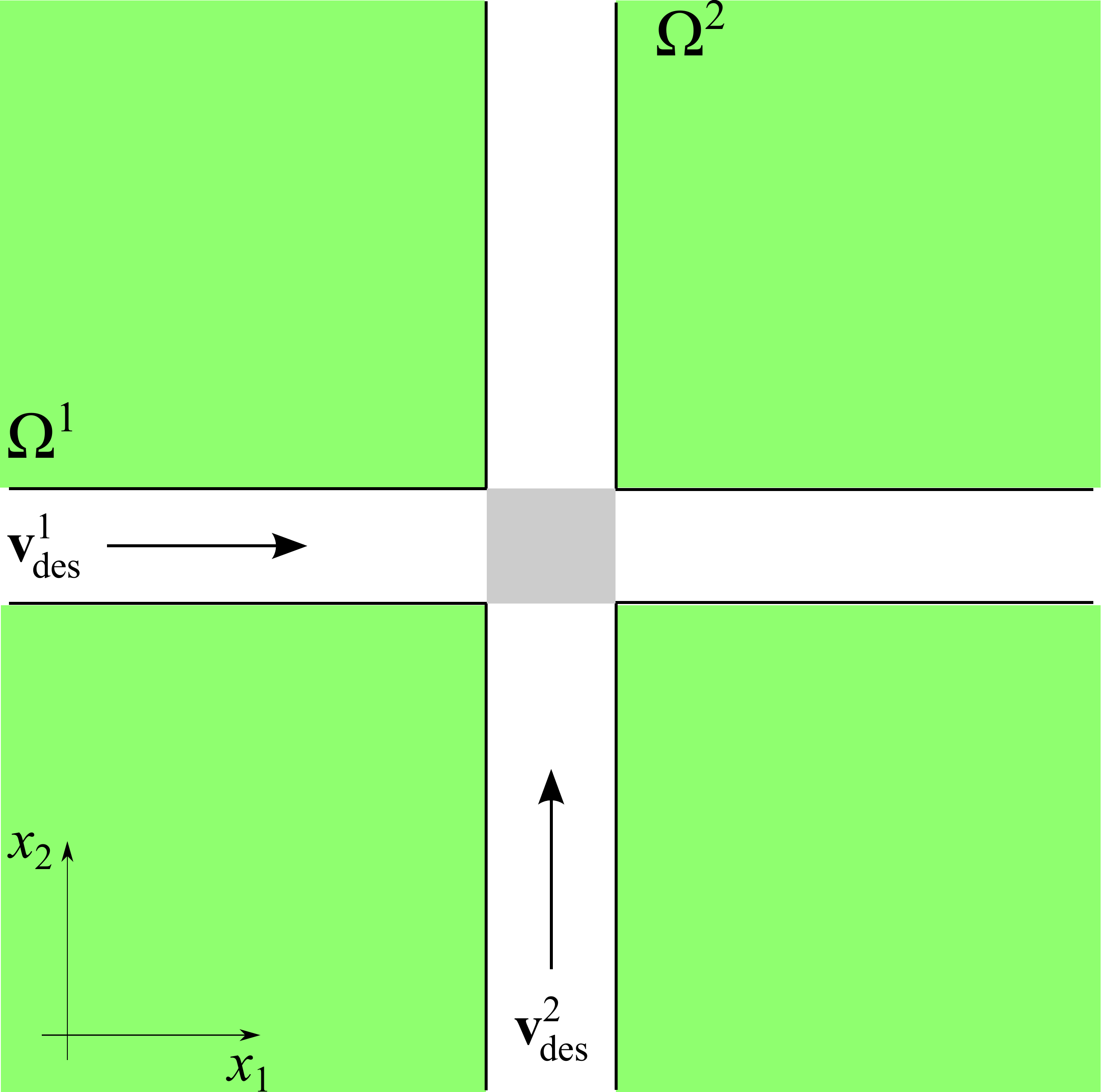}
\caption{Two one-way roads with a junction (shaded)}
\label{fig:domain}
\end{figure}

Let $\Omega^{p}\subset\R^2$, $p=1,\,2$, be the road taken by the $p$-th population of cars, agreeing that $p=1$ labels the horizontal road and $p=2$ the vertical one, see Fig.~\ref{fig:domain}. Then the domain of the problem is $\Omega:=\Omega^{1}\cup\Omega^{2}$, particularly the junction is the subset $\Omega^{1}\cap\Omega^{2}$. Moreover, let $\vec{X}^{i,p}_t\in\Omega^p$ be the position at time $t$ of the $i$-th car of the $p$-th population, that we assume to be ruled by the following equation:
\begin{equation}
	\dot{\vec{X}}^{i,p}_t=\vd^p(\vec{X}^{i,p}_t)+\sum_{q=1}^{2}\sum_{j=1}^{N^q}\vec{K}^{pq}(\vec{X}^{i,p}_t,\,\vec{X}^{j,q}_t),
	\label{eq:micro_model}
\end{equation}
where, from left to right:
\begin{itemize}
\item $\vd^p:\Omega\to\R^2$ is the \emph{desired velocity} of the $p$-th population, namely the one a car would choose if it were alone along the road. For simplicity, we assume that this velocity is constant in time, and, in particular, that $\vd^1$ is rightward and $\vd^2$ upward as indicated in Fig.~\ref{fig:domain};
\item $N^q$, $q=1,\,2$, is the total number of cars in the $q$-th population;
\item $\vec{K}^{pq}:\Omega\times\Omega\to\R^2$ is the \emph{interaction kernel}, that is a mapping modeling how the presence of cars of the $q$-th population may induce the $i$-th car of the $p$-th population to modify its desired velocity. When $q=p$ interactions are \emph{endogenous}, otherwise they are \emph{exogenous}.
\end{itemize}

Model \eqref{eq:micro_model} is \emph{agent-based}, because it takes invariably a Lagrangian viewpoint: cars are tracked one-by-one in space as time goes by. The way in which we now pass to a more general Eulerian model is by looking at \emph{car distributions} in space rather than at cars themselves. For this, we formally consider the $\vec{X}^{i,p}_t$'s as random variables and we denote by $\mu^p_t$ their probability distribution in $\R^2$. In practice, for all measurable subset $E\subseteq\Omega$, we have:
$$ \mu^p_t(E)=\operatorname{Prob}(\vec{X}^{i,p}_t\in E). $$
The fact that $\mu^p_t$ is not labeled by the index $i$ translates the simplifying assumption that cars are \emph{indistinguishable} within the same population, so that $\mu^p_t$ is the common law of all of the $\vec{X}^{i,p}_t$'s, $i=1,\,\dots,\,N^p$.

Using stochastic It\^o's calculus and ensemble averages, we get from \eqref{eq:micro_model} the following equation for the measure $\mu^p_t$ (see \cite{scianna2011dcm} for technical details):
\begin{equation}
	\frac{\partial\mu^p_t}{\partial t}+\nabla\cdot{(\mu^p_t\vec{v}^p_t)}=0
	\label{eq:measure}
\end{equation}
with
\begin{equation}
	\vec{v}^p_t(\vec{x})=\vd^p(\vec{x})+\sum_{q=1}^{2}N^q\intl_{\Omega}\vec{K}^{pq}(\vec{x},\,\vec{y})\,d\hat{\mu}^q_t(\vec{y}\vert\vec{x}),
	\label{eq:velocity}
\end{equation}
where $\hat{\mu}^q_t(\cdot\vert\vec{x})$ is the probability law of the $q$-th population conditioned to the point $\vec{x}$.

In order to close model \eqref{eq:measure}-\eqref{eq:velocity} we may assume $\hat{\mu}^q_t(\cdot\vert\vec{x})=\mu^q_t(\cdot)$. In this case, it is interesting to note that the empirical measure
\begin{equation}
	\mu^p_t=\frac{1}{N^p}\sum_{i=1}^{N^p}\delta_{\vec{X}^{i,p}_t} \qquad (\delta=\text{Dirac delta})
	\label{eq:empirical.meas}
\end{equation}
is a solution to \eqref{eq:measure}-\eqref{eq:velocity} if and only if the $\vec{X}^{i,p}_t$'s solve \eqref{eq:micro_model}. Hence model \eqref{eq:micro_model} is contained into the mathematical structure \eqref{eq:measure}-\eqref{eq:velocity}. However the latter is much richer, indeed it also allows us to choose for $\mu^p_t$ a continuous distribution:
\begin{equation}
	\mu^p_t=\rho^p_t\mathcal{L}^2,
	\label{eq:continuous.meas}
\end{equation}
where $\mathcal{L}^2$ is the Lebesgue measure in $\R^2$ (viz. the area measure) and $\rho^p_t=\rho^p_t(\vec{x}):\Omega\to[0,\,+\infty)$ is the density distribution function in space of cars of the $p$-th population.

Describing the distribution of vehicles via measure \eqref{eq:empirical.meas} corresponds to modeling cars as geometrical points with mass structure, thus emphasizing the intrinsic \emph{granularity} of their distribution. Conversely, using measure \eqref{eq:continuous.meas} corresponds to modeling cars as massless points distributed in space, thus focusing mainly on their continuous stream. If we plug separately \eqref{eq:empirical.meas} and \eqref{eq:continuous.meas} into \eqref{eq:measure}, we get the following dual microscopic/macroscopic representation (see \cite{cristiani2011mmg} for technical details):
\begin{equation}
	\begin{cases}
		\dot{\vec{X}}^{i,p}_t=\vec{v}^p_t(\vec{X}^{i,p}_t) \\[2mm]
		\dfrac{\partial\rho^p_t}{\partial t}+\nabla\cdot{(\rho^p_t\vec{v}^p_t)}=0
	\end{cases}
	\label{eq:dual}
\end{equation}
where $\vec{v}^p_t$ is given by \eqref{eq:velocity}. The two dynamics can be profitably coupled, toward a multiscale model, by the following choice of the conditioned measure:
\begin{equation}
	\hat{\mu}^q_t=\theta\frac{1}{N^q}\sum_{j=1}^{N^q}\delta_{\vec{X}^{j,q}_t}+(1-\theta)\rho^q_t\mathcal{L}^2,
		\qquad q=1,\,2,
	\label{eq:multiscale.meas}
\end{equation}
where $\theta\in[0,\,1]$ is a parameter fixing how much the coupling is biased toward the microscopic or the macroscopic description. The 
advection velocity resulting by inserting \eqref{eq:multiscale.meas} into \eqref{eq:velocity} is:
\begin{multline*}
	\vec{v}^p_t(\vec{x})=\vd^p(\vec{x})+\sum_{q=1}^{2}N^q
		\Biggl(\theta\frac{1}{N^q}\sum_{j=1}^{N^q}\vec{K}^{pq}(\vec{x},\,\vec{X}^{j,q}_t) \\
	+(1-\theta)\intl_{\Omega}\vec{K}^{pq}(\vec{x},\,\vec{y})\rho^q_t(\vec{y})\,d\vec{y}\Biggr),
\end{multline*}
which practically realizes the coupling of the microscopic and macroscopic scale in \eqref{eq:dual}. Notice that $\theta=0$ corresponds to a genuinely macroscopic coupling, with the mass points $\vec{X}^{i,p}_t$ passively advected by the continuous flow. Conversely, $\theta=1$ corresponds to a genuinely microscopic coupling, with the density $\rho^p_t$ passively advected by the velocity field generated by the points $\vec{X}^{i,p}_t$. Finally, $0<\theta<1$ corresponds to a multiscale coupling, in which both the mass points $\vec{X}^{i,p}_t$ and the density $\rho^p_t$ contribute actively to the resulting advection velocity.

\subsection{Car interactions}
The multiscale model consisting of \eqref{eq:velocity}, \eqref{eq:dual}, and \eqref{eq:multiscale.meas} still needs the interaction kernel $\vec{K}^{pq}$ to be specified. Inspired by the classical one-dimensional microscopic \emph{follow-the-leader} model (see e.g., \cite{aw2002dct}):
$$ \ddot{x}_i=\alpha\frac{\dot{x}_{i+1}-\dot{x}_i}{{\left(x_{i+1}-x_i\right)}^{1+\gamma}} \qquad (\alpha,\,\gamma>0), $$
which integrated in time yields
$$ \dot{x}_i=C-\frac{\eta}{{\left(x_{i+1}-x_i\right)}^{\gamma}} \qquad (\eta:=\alpha/\gamma) $$
for a suitable constant $C$ (to be possibly regarded as the one-dimensional counterpart of the desired velocity), we generalize it to the present two-dimensional framework by setting:
\begin{equation}
	\vec{K}^{pq}(\vec{x},\,\vec{y})=-\min\left\{\frac{\eta^{pq}}{\abs{\vec{y}-\vec{x}}^\gamma},\,M^{pq}\right\}
		\ind{S_{R^{pq}}(\vec{x})}(\vec{y})\hat{\vec{r}}(\vec{x},\,\vec{y}),
	\label{eq:int.kernel}
\end{equation}
where, from left to right:
\begin{itemize}
\item $\eta^{pq}>0$ is the \emph{interaction rate}, namely a parameter measuring the intensity of the interaction between cars of the $p$-th and $q$-th population;
\item $M^{pq}>0$ is a \emph{sensitivity threshold} necessary for avoiding the singularity when $\vec{y}$ approaches $\vec{x}$ (remember that cars are modeled, in any case, as dimensionless points);
\item $\ind{E}$ is, for a given set $E$, the indicator function of $E$ (i.e., $\ind{E}(\vec{x})=1$ if $\vec{x}\in E$, $0$ otherwise);
\item $S_{R^{pq}}(\vec{x})\subset\Omega$ is the \emph{interaction neighborhood}, i.e., a subset of the ball centered at $\vec{x}$ with radius $R^{pq}>0$, where the presence of neighboring cars actually perturbs the car located in $\vec{x}$;
\item $\hat{\vec{r}}(\vec{x},\,\vec{y}):=(\vec{y}-\vec{x})/\abs{\vec{y}-\vec{x}}$ is the unit vector pointing in the direction of the relative position of the interacting cars.
\end{itemize}

We remark that the interaction neighborhood makes car interactions \emph{nonlocal} in space; furthermore, whenever it does not coincide with the whole ball, it makes them also \emph{anisotropic}. Coherently with the observation that cars are indeed anisotropic particles \cite{daganzo1995rso}, we model the endogenous ($p=q$) interaction neighborhood as the right half-ball in the road $\Omega^1$ and the upper half-ball in the road $\Omega^2$, for translating the fact that drivers normally look ahead only. Conversely, we model the exogenous ($p\ne q$) interaction neighborhood as the intersection of the right and bottom half-balls in the road $\Omega^1$ and the upper and left half-balls in the road $\Omega^2$, for translating the fact that drivers look also in the direction of the incoming opposite flow of cars, especially at the junction.

\subsection{Coupling 2D microscopic and 1D macroscopic dynamics}
Model \eqref{eq:velocity}-\eqref{eq:dual}-\eqref{eq:multiscale.meas}-\eqref{eq:int.kernel} is fully two-dimensional. Nevertheless a two-dimensional description, while being possibly relevant at the microscopic scale for capturing the detail of lateral displacements of single cars, is actually less necessary at the macroscopic scale, where the interest is mainly in the propagation of density waves along the road length. For such a purpose, a one-dimensional description is clearly enough.

Our measure-theoretic setting allows us to obtain easily and rigorously such a multiscale multidimensional coupling. It suffices to assume that the density $\rho^p_t=\rho^p_t(\vec{x})$, $\vec{x}=(x_1,\,x_2)$, depends actually on one space variable only, particularly the one corresponding to the longitudinal direction in either road:
$$
	\begin{array}{rcll}
		\dfrac{\partial\rho^1_t}{\partial x_2}=0 & \Rightarrow & \rho^1_t=\rho^1_t(x_1) & \text{in\ } \Omega^1 \\[3mm]
		\dfrac{\partial\rho^2_t}{\partial x_1}=0 & \Rightarrow & \rho^2_t=\rho^2_t(x_2) & \text{in\ } \Omega^2.
	\end{array}
$$
The mathematical model remains formally the same, but for the fact that the advection velocity $\vec{v}^p_t$ in the equation for the density, cf. \eqref{eq:dual}, has to be projected onto the longitudinal direction of either road. Specifically, if $\hat{\vec{u}}^p$ denotes the unit vector in the longitudinal direction of $\Omega^p$ (i.e., the horizontal unit vector $\vec{i}$ in $\Omega^1$ and the vertical unit vector $\vec{j}$ in $\Omega^2$), the 2D microscopic-1D macroscopic model reads:
\begin{equation}
	\begin{cases}
		\dot{\vec{X}}^{i,p}_t=\vec{v}^p_t(\vec{X}^{i,p}_t) \\[2mm]
		\dfrac{\partial\rho^p_t}{\partial t}+\dfrac{\partial}{\partial x_p}{(\rho^p_t\vec{v}^p_t\cdot\hat{\vec{u}}^p)}=0,
	\end{cases}
	\label{eq:1D.macro-2D.micro}
\end{equation}
where $\cdot$ denotes the standard inner product in $\R^2$.

Notice that, thanks to the fact that $\vec{v}^p_t$ is projected in one dimension after being computed as genuinely two-dimensional, multiscale interactions are automatically evaluated with the correct dimensionality. The microscopic two-dimensional points $\vec{X}^{i,p}_t$ perceive a spatially two-dimensional density, simply constant in the transverse direction. On the other hand, the macroscopic one-dimensional density $\rho^p_t$ is affected only by the longitudinal component of the interactions with the mass points, which are however two-dimensional.

\subsection{Variable multiscale coupling}
\label{sec:variable.theta}
The parameter $\theta$, which determines the multiscale coupling, is assumed to be constant in \eqref{eq:multiscale.meas}. This implies, in general, that the microscopic and macroscopic scales contribute jointly to the dynamics everywhere in the domain $\Omega$. However, one may argue that their specific contributions are actually relevant in different sub-domains. For instance, it might be important to retain the microscopic detail at the junction, where the actual granularity of the flow of cars plays a major role in shaping the exogenous interactions between the crossing populations. Conversely, away from the junction, where only endogenous interactions occur, a gross continuous description may be appropriate.

This amounts to converting $\theta$ into a function $\theta(\vec{x}):\Omega\to[0,\,1]$, so that the conditioned measure depends now explicitly on the conditioning point $\vec{x}$:
\begin{equation}
	\hat{\mu}^q_t(\cdot\vert\vec{x})=
		\theta(\vec{x})\frac{1}{N^q}\sum_{j=1}^{N^q}\delta_{\vec{X}^{j,q}_t}+(1-\theta(\vec{x}))\rho^q_t\mathcal{L}^2.
		\label{eq:multiscale.meas.theta.var}
\end{equation}
It is trivial to check that \eqref{eq:multiscale.meas.theta.var} is indeed a probability measure on $\R^2$ for each $\vec{x}\in\Omega$.

The scale switch at the junction described above can be obtained, for instance, by means of the following function:
\begin{equation}
	\theta(\vec{x})=\ind{V}(\vec{x})=
		\begin{cases}
			1 & \text{if\ } \vec{x}\in V \\
			0 & \text{if\ } \vec{x}\in \Omega\setminus V,
		\end{cases}
\end{equation}
where $V\subset\R^2$ is a neighborhood of the junction $\Omega^1\cap\Omega^2$ (see Fig.~\ref{fig:domain.variable.theta}).

\section{Numerical simulations}
In this section we present some numerical simulations that show the potential of the multiscale multidimensional model described above. The numerical method used for solving \eqref{eq:1D.macro-2D.micro}, along with the multiscale algorithm, can be easily derived from those described in \cite{cristiani2011mmg}. The main difference here is that the equations for the densities are one-dimensional, therefore the two-dimensional quantities computed by the microscopic part of the algorithm have to be projected to scalar quantities before they can be used by the macroscopic part.

The two roads forming the computational domain are the stripes $\Omega^1=[0,\,200\unit{m}]\times[95\unit{m},\,105\unit{m}]$ and $\Omega^2=[95\unit{m},\,105\unit{m}]\times[0,\,200\unit{m}]$. Microscopic cars can freely move in the whole domain, while car densities are constrained to a numerical grid consisting of $N$ nodes in each road. The space step is $\Delta x=200/N$. Desired velocities are $\vd^1=(10\unit{m/s})\vec{i}$ and $\vd^2=(10\unit{m/s})\vec{j}$. The exponent $\gamma$ in the expression of the interaction kernel $\vec{K}^{pq}$ is fixed to $\gamma=1$.

Inflow boundary conditions are prescribed as follows.
\begin{itemize}
\item Microscopic cars of either population enter the domain from $x_1=0$ and $x_2=0$, respectively, every $\Delta{t}_b=0.9\unit{s}$, randomly distributed along the width of the roads.
\item Car densities are imposed for all $t\geq 0$ at the same boundaries: $\rho_t^p(x_p=0)=\rho_b$, where the constant $\rho_b$ is computed in such a way that the total macroscopic mass flowing into the domain up to the final time $\Tmax>0$ of the simulation equals the total number of microscopic cars. Recalling that the $\rho^p_t$'s are probability densities, this is written as:
$$ \intl_{0}^{\Tmax}\intl_{\textup{road width}}{(N^p\rho^p_t\vd^p\cdot\hat{\vec{u}}^p)}_{\vert x_p=0}\,dx\,dt=N^p, $$
whence, noting that $N^p\sim\Tmax/\Delta{t}_b$,
$$ \rho_b=\frac{1}{N^p\abs{\vd^p}\Delta{t}_b\ell} \qquad (\ell>0\ \text{road width}). $$
\end{itemize}
At the end of either road ($x_p=200\unit{m}$) we assume that cars can freely exit with their desired velocity. At the macroscopic scale, this is equivalent to imagine $\rho^p_t(x_p)=0$ for $x_p>200\unit{m}$ and $t>0$.

Note that describing the roads as one-dimensional domains (rather than two-dimensional) allows one to avoid boundary conditions along the lateral edges, which are difficult to treat analytically and can affect heavily the solution.

Finally, the (variable) time step is chosen as:
$$ \Delta{t}=\min\{0.05\unit{s},\,\Delta{t}_{CFL}\}, $$
where $\Delta{t}_{CFL}$ is the time step imposed by the CFL condition at the macroscopic scale: 
$$ \Delta{t}_{CFL}\max_{\substack{p=1,\,2 \\ x_p\in[0,\,200\unit{m}]}}\abs{(\vec{v}^p_t\cdot\hat{\vec{u}}^p)(x_p)}=\Delta{x}. $$

\begin{remark}
The above parameters, as well as those that will be specified in the subsequent numerical tests, are intended to be realistic but exploratory. In particular, they do not stem from any real measurement.
\end{remark}

\begin{remark}
As recalled in the Introduction, the simulated junction is not regulated by any give-way rule nor by red-green traffic light cycles.
\end{remark}

\subsection{Test 1: Inducing the traffic light effect in the density}
In this test we aim at showing how granularity triggers self-organization, and, in particular, which effect it has on the macroscopic dynamics. Here we chose $N=100$, $\eta^{11}=\eta^{22}=1\unit{m$^2$/s}$, $\eta^{12}=\eta^{21}=35\unit{m$^2$/s}$, $R^{11}=R^{22}=10\unit{m}$, $R^{12}=R^{21}=20\unit{m}$, $M^{11}=M^{22}=15\unit{m/s}$, $M^{12}=M^{21}=50\unit{m/s}$.

\begin{figure}[!t]
\centering
\includegraphics[scale=0.6]{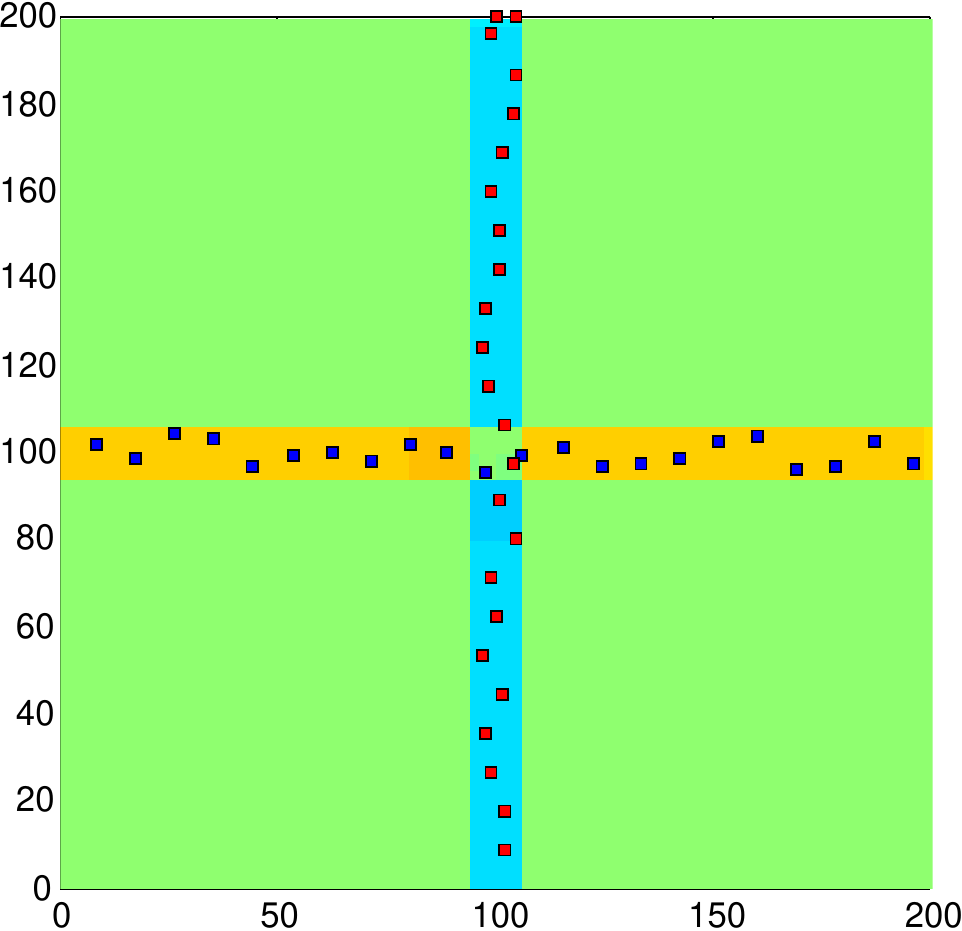}
\caption{Test 1: simulation with $\theta=0$ (genuinely macroscopic dynamics). No particular effect is visible at the junction}
\label{fig:T1_macro}
\end{figure}

Figure~\ref{fig:T1_macro} shows the outcome of the simulation with $\theta=0$, i.e., with the macroscopic scale leading the dynamics of both point cars and densities. The latter flow quite uniformly and do not show any notable behavior at the junction. 

\begin{figure}[!t]
\centering
\includegraphics[scale=0.6]{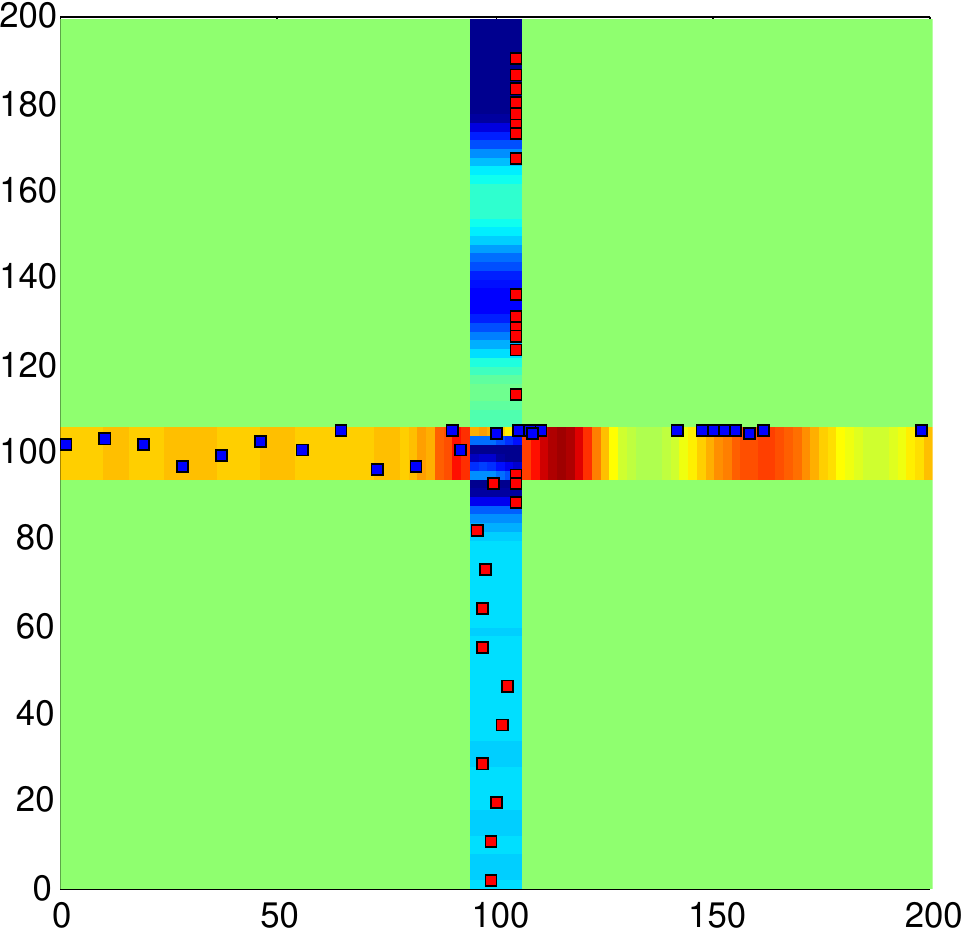}
\caption{Test 1: simulation with $\theta=0.7$ (multiscale dynamics). Traffic light effect beyond the junction is visible in both densities and microscopic cars}
\label{fig:T1_micromacro}
\end{figure}

Figure \ref{fig:T1_micromacro} shows instead the outcome of the same simulation with $\theta=0.7$. Here a clear traffic light effect appears beyond the junction. This is due to the fact that interactions among point cars are much different from those among densities: more granularity expedites the break of symmetry and triggers the typical alternate occupation of the space also at the macroscopic scale.

\begin{figure}[!t]
\centering
\includegraphics[scale=0.2]{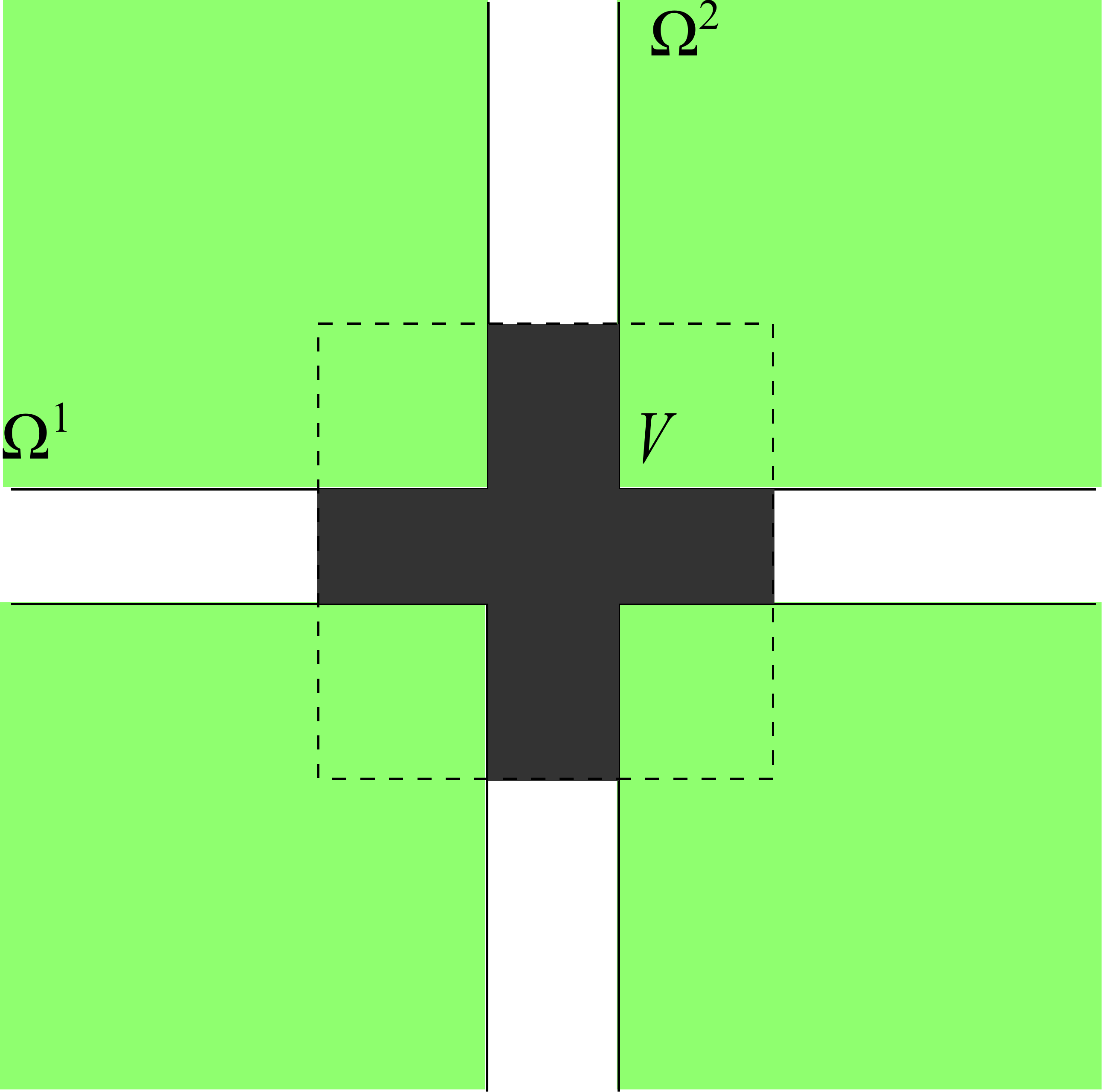}
\caption{Computational domain partitioned in two regions: the microscopic scale leads the dynamics in the black area $V$, the macroscopic scale in the remaining region $(\Omega^1\cup\Omega^2)\setminus V$}
\label{fig:domain.variable.theta}
\end{figure}

\subsection{Test 2: Space-dependent $\theta$}
In this test we assume the scale parameter $\theta$ to depend on $\vec{x}$, as described in Sect.~\ref{sec:variable.theta}. In more detail, we choose $\theta=1$ inside a neighborhood $V$ of the junction $\Omega^1\cap\Omega^2$ centered at $(100\unit{m},\,100\unit{m})$ and ideally contained in a square of edge length $40\unit{m}$, and $\theta=0$ elsewhere (see Fig.~\ref{fig:domain.variable.theta}). This is a natural choice since we expect that the granularity plays a role mainly at the junction. The other parameters are $N=200$, $\eta^{11}=\eta^{22}=7\unit{m$^2$/s}$, $\eta^{12}=\eta^{21}=27\unit{m$^2$/s}$, $R^{11}=R^{22}=5\unit{m}$, $R^{12}=R^{21}=10\unit{m}$, $M^{11}=M^{22}=20\unit{m/s}$, $M^{12}=M^{22}=40\unit{m/s}$.

\begin{figure}[!t]
\centering
\includegraphics[scale=0.6]{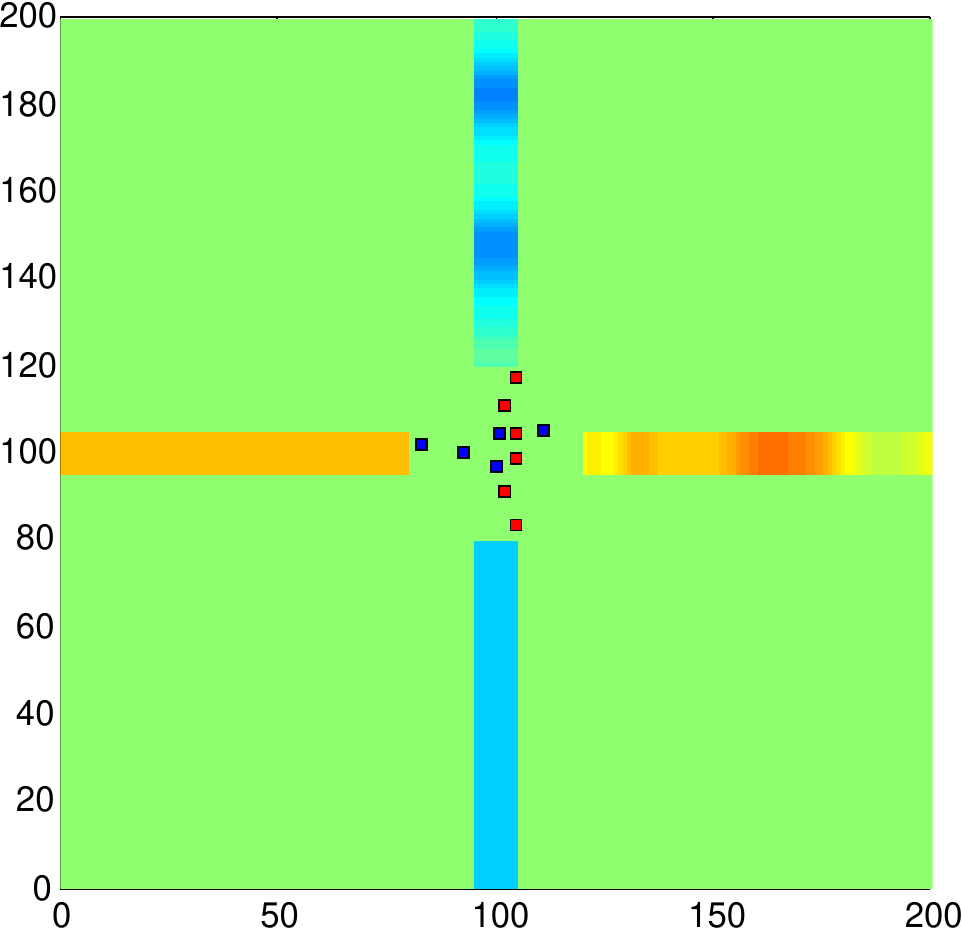}
\caption{Test 2: $\theta$ depends on $\vec{x}$. The microscopic scale leads near the junction, the macroscopic scale elsewhere}
\label{fig:T2_plot2D_mixed}
\end{figure}

Figure~\ref{fig:T2_plot2D_mixed} shows the outcome of the simulation at time $t=23\unit{s}$. Only the leading scale is plotted, namely point cars near the junction and densities elsewhere, but we recall that \emph{both scales are present everywhere and evolve at all times, thus no transmission conditions are needed at the interfaces of} $V$. Microscopic interactions trigger a self-organizing alternate passage beyond the junction, perfectly visible in terms of density waves, which persists also when the microscopic scale ceases its influence on the dynamics. This test shows that self-organization can be obtained by adding granularity only at the junction rather than from the very beginning like in Test 1 (cf. Fig.~\ref{fig:T1_micromacro}).

\begin{figure}[!t]
\centering
\includegraphics[scale=0.6]{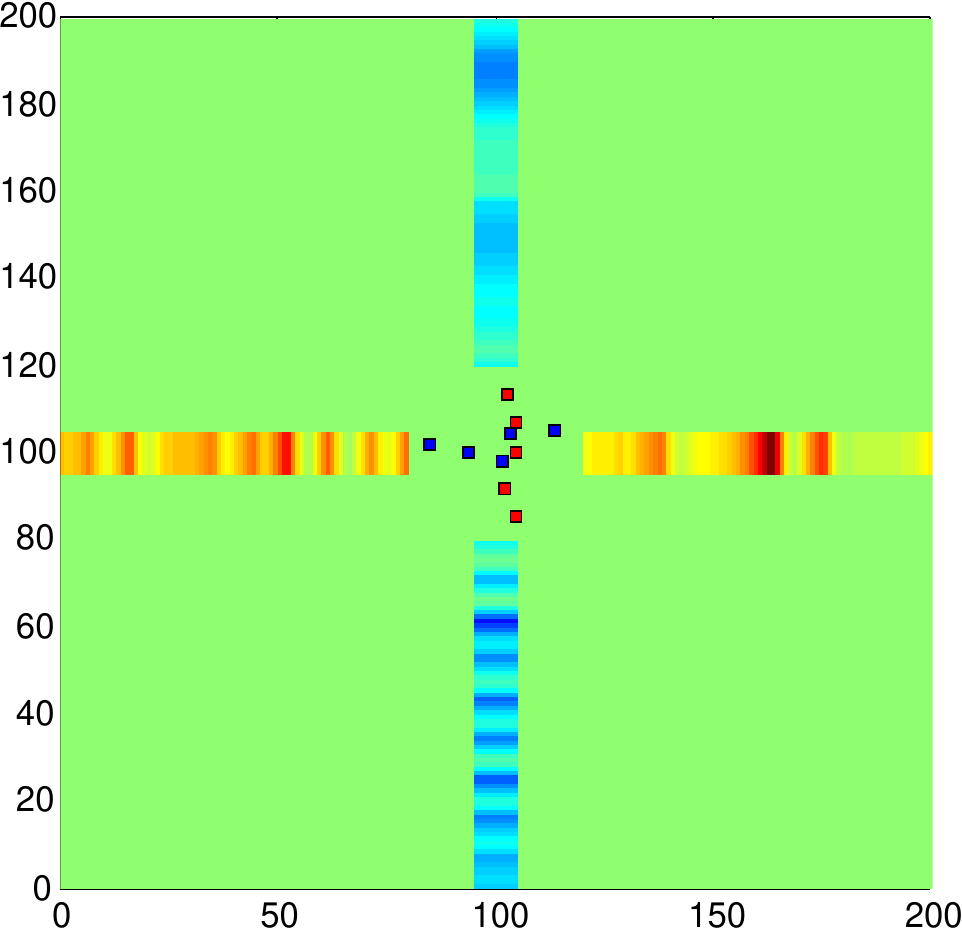}
\caption{Test 2: $\theta(\vec{x})\equiv 1$: the microscopic scale leads everywhere}
\label{fig:T2_plot2D_micro}
\end{figure}

\begin{figure}[!t]
\centering
\includegraphics[scale=0.6]{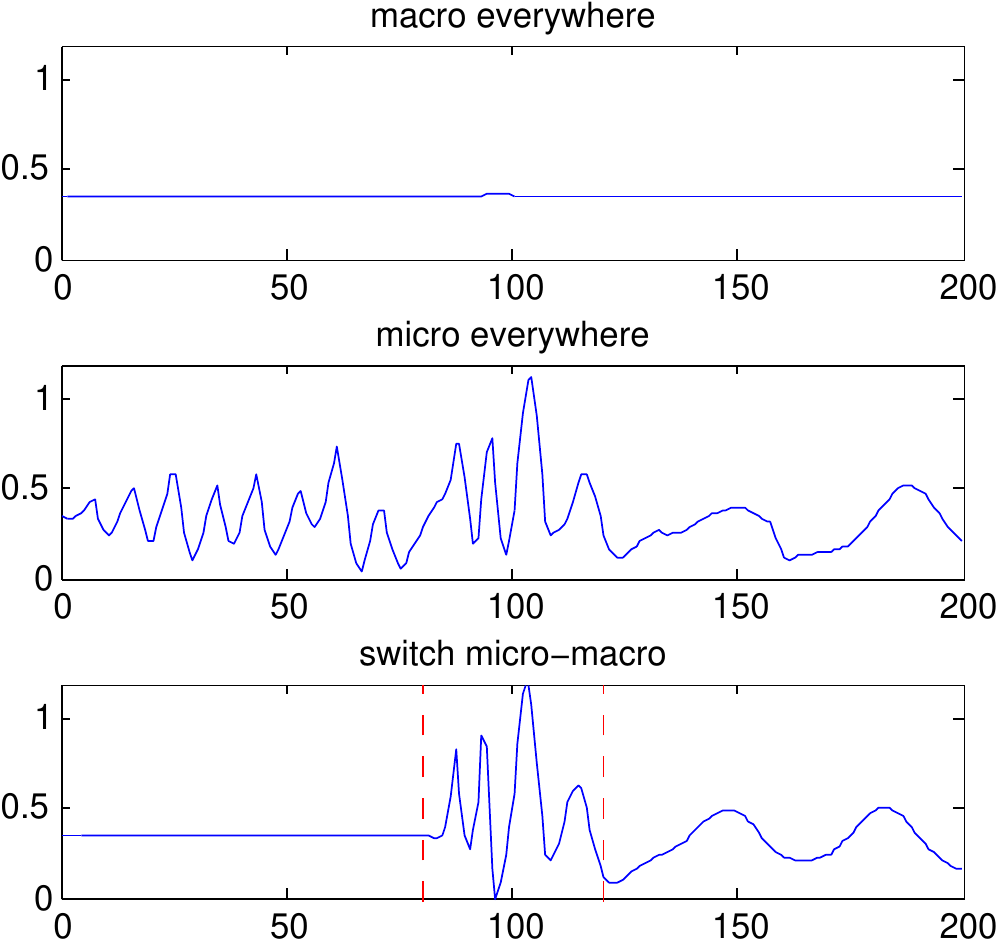}
\caption{Test 2: Car density along the vertical road at time $t=23\unit{s}$ with $\theta(\vec{x})\equiv 0$ (top), $\theta(\vec{x})\equiv 1$ (middle), and $\theta(\vec{x})=\ind{V}(\vec{x})$ (bottom)}
\label{fig:T2_plot1D}
\end{figure}

It is interesting to compare the previous result with the extreme cases $\theta(\vec{x})\equiv 0$ and $\theta(\vec{x})\equiv 1$. The simulation with vanishing $\theta$ does not show any particular effect at the junction, analogously to the result depicted in Fig.~\ref{fig:T1_macro}. Conversely, with $\theta(\vec{x})\equiv 1$ (dynamics fully led by the microscopic scale) the traffic light effect comes up at and beyond the junction, see Fig.~\ref{fig:T2_plot2D_micro}. However, an annoying effect also appears before and, less visible, beyond the junction: behind every point car (not shown) density accumulates, due to strong interactions with Dirac delta's. Such an effect perturbs the density, especially where it should be ideally constant (e.g., before the junction), see also the same effect in Fig. 6.1(c) of \cite{cristiani2011mmg}. \emph{This motivates the use of the microscopic scale only where it is really needed}. To better quantify this qualitative discussion, we plotted in Fig.~\ref{fig:T2_plot1D} the one-dimensional density along the vertical road. Fully macroscopic dynamics do not show any appreciable effect either at the junction or beyond it. Conversely, fully microscopic dynamics generate irregular waves beyond the junction, as well as a noisy density profile before it. Finally, mixed dynamics (fully microscopic near the junction, fully macroscopic elsewhere) produce a constant density before the junction and clear density wave packets beyond it.

\bibliographystyle{plain}
\bibliography{CePbTa-crossroads-CDC}

\begin{thebibliography}{10}

\bibitem{aw2002dct}
A.~Aw, A.~Klar, T.~Materne, and M.~Rascle.
\newblock Derivation of continuum traffic flow models from microscopic
  follow-the-leader models.
\newblock {\em SIAM J. Appl. Math.}, 63(1):259--278, 2002.

\bibitem{aw2000rso}
A.~Aw and M.~Rascle.
\newblock Resurrection of ``second order'' models of traffic flow.
\newblock {\em SIAM J. Appl. Math.}, 60(3):916--938, 2000.

\bibitem{bellomo2011mtc}
N.~Bellomo and C.~Dogb{\'e}.
\newblock On the modelling of traffic and crowds. {A} survey of models,
  speculations, and perspectives.
\newblock {\em SIAM Rev.}, 53(3):409--463, 2011.

\bibitem{chakroborty2004mmd}
P.~Chakroborty, S.~Agrawal, and K.~Vasishtha.
\newblock Microscopic modeling of driver behavior in uninterrupted traffic
  flow.
\newblock {\em J. Transp. Eng.}, 130(4):438--451, 2004.

\bibitem{cristiani2011mmg}
E.~Cristiani, B.~Piccoli, and A.~Tosin.
\newblock Multiscale modeling of granular flows with application to crowd
  dynamics.
\newblock {\em Multiscale Model. Simul.}, 9(1):155--182, 2011.

\bibitem{daganzo1995rso}
C.~F. Daganzo.
\newblock Requiem for second-order fluid approximation of traffic flow.
\newblock {\em Transportation Res.}, 29(4):277--286, 1995.

\bibitem{garavello2006tfn}
M.~Garavello and B.~Piccoli.
\newblock {\em Traffic flow on networks}, volume~1.
\newblock American Institute of Mathematical Sciences, Springfield, MO, USA,
  2006.

\bibitem{gazis1961nfl}
D.~C. Gazis, R.~Herman, and R.~W. Rothery.
\newblock Nonlinear follow-the-leader models of traffic flow.
\newblock {\em Operations Res.}, 9:545--567, 1961.

\bibitem{helbing2001trs}
D.~Helbing.
\newblock Traffic and related self-driven many-particle systems.
\newblock {\em Rev. Modern Phys.}, 73(4):1067--1141, 2001.

\bibitem{helbing2002mms}
D.~Helbing, A.~Hennecke, V.~Shvetsov, and M.~Treiber.
\newblock Micro- and macro-simulation of freeway traffic.
\newblock {\em Math. Comput. Modelling}, 35(5--6):517--547, 2002.

\bibitem{helbing2007son}
D.~Helbing, J.~Sigmeier, and S.~L\"{a}mmer.
\newblock Self-organized network flows.
\newblock {\em Netw. Heterog. Media}, 2(2):193--210, 2007.

\bibitem{herty2009mky}
M.~Herty and S.~Moutari.
\newblock A macro-kinetic hybrid model for traffic flow on road networks.
\newblock {\em Comput. Methods Appl. Math.}, 9(3):238--252, 2009.

\bibitem{hoogendoorn2001soa}
S.~P. Hoogendoorn and P.~H.~L. Bovy.
\newblock State-of-the-art of vehicular traffic flow modelling.
\newblock {\em J. Syst. Cont. Eng.}, 215(4):283--303, 2001.

\bibitem{kerner2002mmp}
B.~S. Kerner and S.~L. Klenov.
\newblock A microscopic model for phase transitions in traffic flow.
\newblock {\em J. Phys. A}, 35(3):L31--L43, 2002.

\bibitem{lattanzio2010cmm}
C.~Lattanzio and B.~Piccoli.
\newblock Coupling of microscopic and macroscopic traffic models at boundaries.
\newblock {\em Math. Models Methods Appl. Sci.}, 20(12):2349--2370, 2010.

\bibitem{lighthill1955kw2}
M.~J. Lighthill and G.~B. Whitham.
\newblock On kinematic waves. {II}. {A} theory of traffic flow on long crowded
  roads.
\newblock {\em Proc. Roy. Soc. London. Ser. A.}, 229:317--345, 1955.

\bibitem{piccoli2009vtr}
B.~Piccoli and A.~Tosin.
\newblock Vehicular traffic: {A} review of continuum mathematical models.
\newblock In R.~A. Meyers, editor, {\em Encyclopedia of Complexity and Systems
  Science}, volume~22, pages 9727--9749. Springer, New York, 2009.

\bibitem{richards1956swh}
P.~I. Richards.
\newblock Shock waves on the highway.
\newblock {\em Operations Res.}, 4:42--51, 1956.

\bibitem{scianna2011dcm}
M.~Scianna, A.~Tosin, and L.~Preziosi.
\newblock From discrete to continuous models of cell colonies: {A}
  measure-theoretic approach.
\newblock Preprint, 2011.

\bibitem{treiber2000cts}
M.~Treiber, A.~Hennecke, and D.~Helbing.
\newblock Congested traffic states in empirical observations and microscopic
  simulations.
\newblock {\em Phys. Rev. E}, 62(2):1805--1824, 2000.

\end{thebibliography}

\end{document}